\documentclass[twocolumn]{USG}

\usepackage[numbers,sort&compress]{natbib}

\makeatletter
\patchcmd{\@maketitle}{\includegraphics[width=26.5mm]{Wiley_logo.eps}}{}{}{\PackageWarning{arxiv}{Wiley logo patch failed}}
\patchcmd{\@maketitle}{\includegraphics[width=56mm]{allergy.eps}}{}{}{\PackageWarning{arxiv}{allergy.eps patch failed}}
\def\@@articletype{}
\historydates{}
\def\oddhead@titlepage@info{}
\def\evenhead@titlepage@info{}
\def\oddfoot@titlepage@info{\hfil{\pagenumfont\thepage}\hfil}
\def\evenfoot@titlepage@info{\oddfoot@titlepage@info}
\def\oddhead@headings@info{\hfil{\pagenumfont\thepage}\hfil}
\def\evenhead@headings@info{\oddhead@headings@info}
\makeatother

\begin{document}

\title{Combined photoluminescence and electrical characterization of valley photovoltaic devices explained with an equivalent circuit}

\author[1]{Daixi Xia}
\author[2]{Abhinav S. Sharma}
\author[3]{Andreas Pusch}
\author[3]{Murad J. Y. Tayebjee}
\author[3]{Michael P. Nielsen}
\author[3]{Fiacre E. Rougieux}
\author[4]{Kyle R. Dorman}
\author[4]{Tetsuya D. Mishima}
\author[4]{Michael B. Santos}
\author[5]{Ian R. Sellers}
\author[3]{Nicholas J. Ekins-Daukes}
\author[1,6]{Jacob J. Krich}

\authormark{}
\titlemark{}

\address[1]{\orgdiv{Department of Physics, }\orgname{University of Ottawa, }\orgaddress{\state{Ottawa, Ontario, }\country{Canada}}}
\address[2]{\orgdiv{Chair in Hybrid Nanosystems, Nanoinstitut, Fakult\"{a}t f\"{u}r Physik, }\orgname{Ludwig-Maximilians-Universit\"{a}t M\"{u}nchen, }\orgaddress{\state{Munich, }\country{Germany}}}
\address[3]{\orgdiv{School of Photovoltaic and Renewable Energy Engineering, }\orgname{University of New South Wales, }\orgaddress{\state{Sydney, New South Wales, }\country{Australia}}}
\address[4]{\orgdiv{Department of Physics \& Astronomy, }\orgname{University of Oklahoma, }\orgaddress{\state{Norman, Oklahoma, }\country{USA}}}
\address[5]{\orgdiv{Department of Electrical Engineering, }\orgname{University at Buffalo, }\orgaddress{\state{Buffalo, New York, }\country{USA}}}
\address[6]{\orgdiv{Nexus for Quantum Technologies, }\orgname{University of Ottawa, }\orgaddress{\state{Ottawa, Ontario, }\country{Canada}}}

\corres{Jacob Krich (\email{jkrich@uottawa.ca})}

\fundingInfo{NSERC CREATE TOP-SET (Award 497981); NSERC (RGPIN-2019-06559); UNSW Scientia Program; Australian Research Council DECRA Fellowship (DE230100382)}

\abstract[ABSTRACT]{We present an equivalent-circuit analysis that simultaneously considers
the photoluminescence and electrical responses of valley photovoltaic
devices. Valley photovoltaics (VPV) are a novel concept for hot-carrier
solar cells, where intervalley scattering of electrons in the absorbing
material helps maintain the conduction-band electron populations at
metastable satellite valleys, which are at energies higher than the
conduction-band minimum. We measure intensity- and voltage-dependent
photoluminescence and electrical signals of a VPV device, which has
previously shown S-shaped current-voltage characteristics and thus
low efficiency. We present an equivalent circuit that captures the
S-shaped current-voltage curves and simultaneously describes our photoluminescence
and electric measurements. The equivalent-circuit model indicates
that a reverse diode causes inefficient extraction of electrons and
thus the S shape. We use a Poisson/drift-diffusion model to demonstrate
that the reverse diode can be either explained by the valley-scattering
process itself or a heterojunction barrier. The reverse diode must
be eliminated before VPV devices can achieve high efficiency.}

\maketitle

\section{Introduction}

\begin{figure}
\centering
\includegraphics[width=\columnwidth]{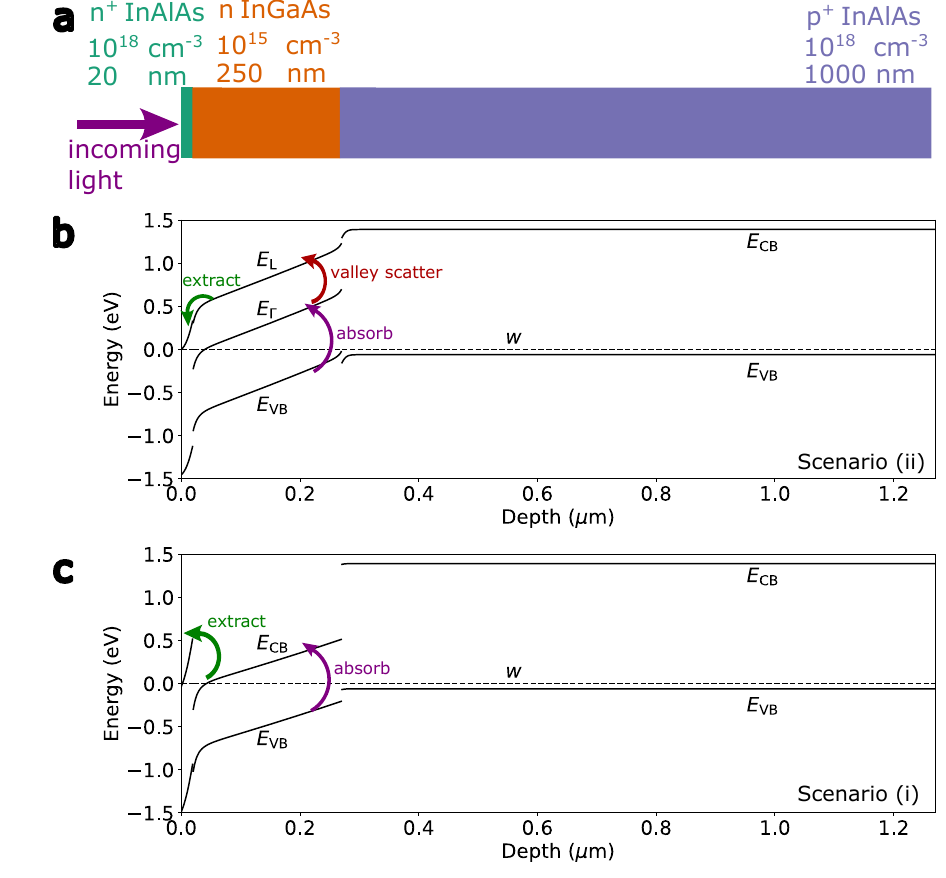}
\caption{\label{fig:device_band_diagrams}(a) Structure of the proof-of-concept
device in Ref.~\cite{esmaielpour_exploiting_2020}. (b) Valley-scattered
simulated band diagram for the VPV device from \cite{esmaielpour_exploiting_2020},
where in the InGaAs region, electrons are generated in the $\Gamma$
valley from photon absorption. These electrons scatter to the L valley
and are extracted with minimal energy barrier to the InAlAs conduction
band. (c) $\Gamma$-extracted simulated band diagram for a case where
no valley scattering is present. The energy barrier for extracting
electrons from InGaAs to InAlAs is much bigger in the $\Gamma$-extracted
case than the L-extracted case. Both band diagrams simulated at equilibrium
using Simudo, our Poisson/drift-diffusion device model\cite{dumitrescu_simudo_2020}.}
\end{figure}

Hot-carrier solar cells have the potential of breaking the Shockley-Queisser
efficiency limit by mitigating thermalization loss \cite{wurfel_particle_2005}.
Hot-carrier solar cells rely on sustaining a high carrier temperature
while extracting carriers at an energy-selective contact \cite{green_third_2003}.
Many efforts have tried to realize high carrier temperatures by exploiting
a phonon bottleneck effect \cite{shah_energy-loss_1985,lugli_nonequilibrium_1987,conibeer_slowing_2008,zhang_review_2022}.
Valley photovoltaics (VPV), using the valley-scattering effect similar
to the Gunn effect, has been proposed as an alternative method to
maintain high carrier temperature \cite{esmaielpour_exploiting_2020,ferry_search_2019}.
The proposed working principle of VPV is that under high electric
field, optically generated carriers are scattered to satellite valleys,
with the population of the L valley exceeding that of the $\Gamma$
valley; extracting the carriers from the L valley increases the device
voltage beyond the limit set by the $\Gamma$-valley bandgap. In Fig.~\ref{fig:device_band_diagrams}a,
we show the nominal device structure of the first VPV device, from
Ref.~\cite{esmaielpour_exploiting_2020}. The device is designed
to have a large built-in electric field of about 25 kV/cm in the middle
InGaAs region, as seen in the simulated band diagram in Fig.~\ref{fig:device_band_diagrams}(b,c).
The conduction band (CB) of InAlAs is designed to energetically align
with the L valley of InGaAs to encourage carrier extraction from the
L valley. Studies have found that electrons are transferred from the
$\Gamma$ to the L valley in this device, but the device suffers from
S-shaped current-voltage curves and thus low efficiency \cite{esmaielpour_exploiting_2020,ferry_search_2019,dorman_toward_2022}.

Ref.~\cite{dorman_toward_2022} suggests that the S shape in VPV
can be attributed to inefficient carrier extraction from the L valley.
The CB minimum of InAlAs is at the $\Gamma$ point, leading to momentum
mismatch at the InAlAs/InGaAs heterojunction (HJ) interface and therefore
inefficient extraction, despite the minimum of the InAlAs CB being
only a few $kT$ away from the minimum of the InGaAs L valley, where
$k$ is the Boltzmann constant and $T$ is device temperature. These
devices can be described either with carrier transport directly from
the L valley or from the $\Gamma$ valley. Fig.~\ref{fig:device_band_diagrams}b
shows the band diagram for the case of ideal carrier extraction, in
which electrons in InGaAs can only be extracted from the L valley;
photo-generated electrons in the $\Gamma$ valley must first scatter
to the L valley before extraction. Alternatively, we can consider
a material where valley scattering is unimportant, and carriers are
extracted directly from the InGaAs $\Gamma$ valley to InAlAs, with
a large energy barrier at the HJ interface, as seen in Fig.~\ref{fig:device_band_diagrams}c.
It is well known that energy barriers, whether at contacts or internal
to the device, can cause S-shaped current-voltage (JV) curves \cite{saive_s-shaped_2019}. 

Equivalent circuit models have been a common tool for analyzing photovoltaic
JV characteristics. Ref.~\cite{garcia-sanchez_lumped_2013} used
an equivalent circuit with anti-parallel diodes to describe S-shaped
JV curves in organic solar cells. Despite the difference in materials,
the S shapes in \cite{garcia-sanchez_lumped_2013} share many similar
characteristics with the JV curves that are observed in VPV, including
reverse saturation, diode-like behavior under large forward bias,
and an S-shape in the fourth quadrant.

In this work, we measure the photoluminescence (PL) and electrical
response of a VPV device nominally identical to the one in Ref.~\cite{esmaielpour_exploiting_2020}.
We find that the model of Ref.~\cite{garcia-sanchez_lumped_2013}
can fit the JV but not the PL measurements. We propose a new equivalent-circuit
model that simultaneously explains PL and electrical measurements
of the VPV device. We give possible physical explanations for the
circuit elements. We use a Poisson/drift-diffusion model to simulate
the VPV devices in two carrier-extraction scenarios and show that
both HJ barrier and valley scattering itself can cause the S-shaped
JV. However, we demonstrate that, as long as our equivalent circuit
describes VPV devices, they will not achieve high efficiency.

\section{Experiments and results}

\begin{figure}
\centering
\includegraphics[width=0.6\columnwidth]{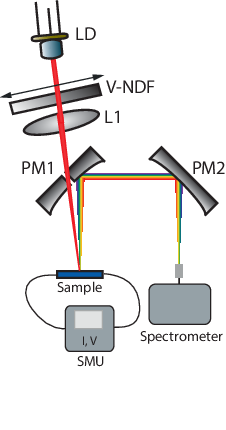}
\caption{\label{fig:experiment} Schematic of photoluminescence measurement
setup. Laser diode (LD) emission (red beam) at 1160 nm was focused
via a lens (L1) to excite the sample. A variable neutral density filter
(V-NDF) was used to control the excitation intensity. The photoluminescence
from the sample (colourful beam) was collimated and focused onto a
spectrometer via parabolic mirrors PM1 and PM2. A source measurement
unit (SMU) was used to electrically bias the sample and also measure
the photocurrent.}
\end{figure}

We study a VPV device with structure shown in Fig.~\ref{fig:device_band_diagrams}a,
nominally identical to that in Ref.~\cite{esmaielpour_exploiting_2020}.
The device has, from top to bottom, 20~nm n$^{+}$-In$_{0.52}$Al$_{0.48}$As,
250 nm p-In$_{0.53}$Ga$_{0.47}$As, 1000 nm p$^{+}$-In$_{0.52}$Al$_{0.48}$As,
on InP substrate. The doping levels are $10^{18}$ cm$^{-3}$, $10^{15}$
cm$^{-3}$, and $10^{18}$ cm$^{-3}$, respectively. Figure~\ref{fig:experiment}
illustrates the PL setup. A laser diode (Innolume GC-1160-90-TO-200-A)
was used to generate the 1160~nm wavelength excitation beam, and
a variable neutral density (ND) filter was used to control the intensity
of the beam, which was focused onto the device. The intensities were
set to be similar to the integrated 1-Sun spectrum (100 mW/cm$^{2}$).
The resultant photoluminescence spectrum was measured via a spectrometer
(BWTek Sol 1.7 TE Cooled InGaAs Array Spectrometer). For the intensity
dependent measurements, the Keithley 2470 source measurement unit
(SMU) was used to hold the sample at short circuit. By adjusting the
position of the variable ND filter, intensity-dependent PL spectra
were measured. For the bias-dependent measurements, the intensity
of the excitation beam was fixed to 136 mW/cm$^{2}$. The SMU was
used to vary the applied bias and measure the photocurrent; the PL
spectrum was concurrently measured at each voltage point.

\begin{figure}
\centering
\includegraphics[width=\columnwidth]{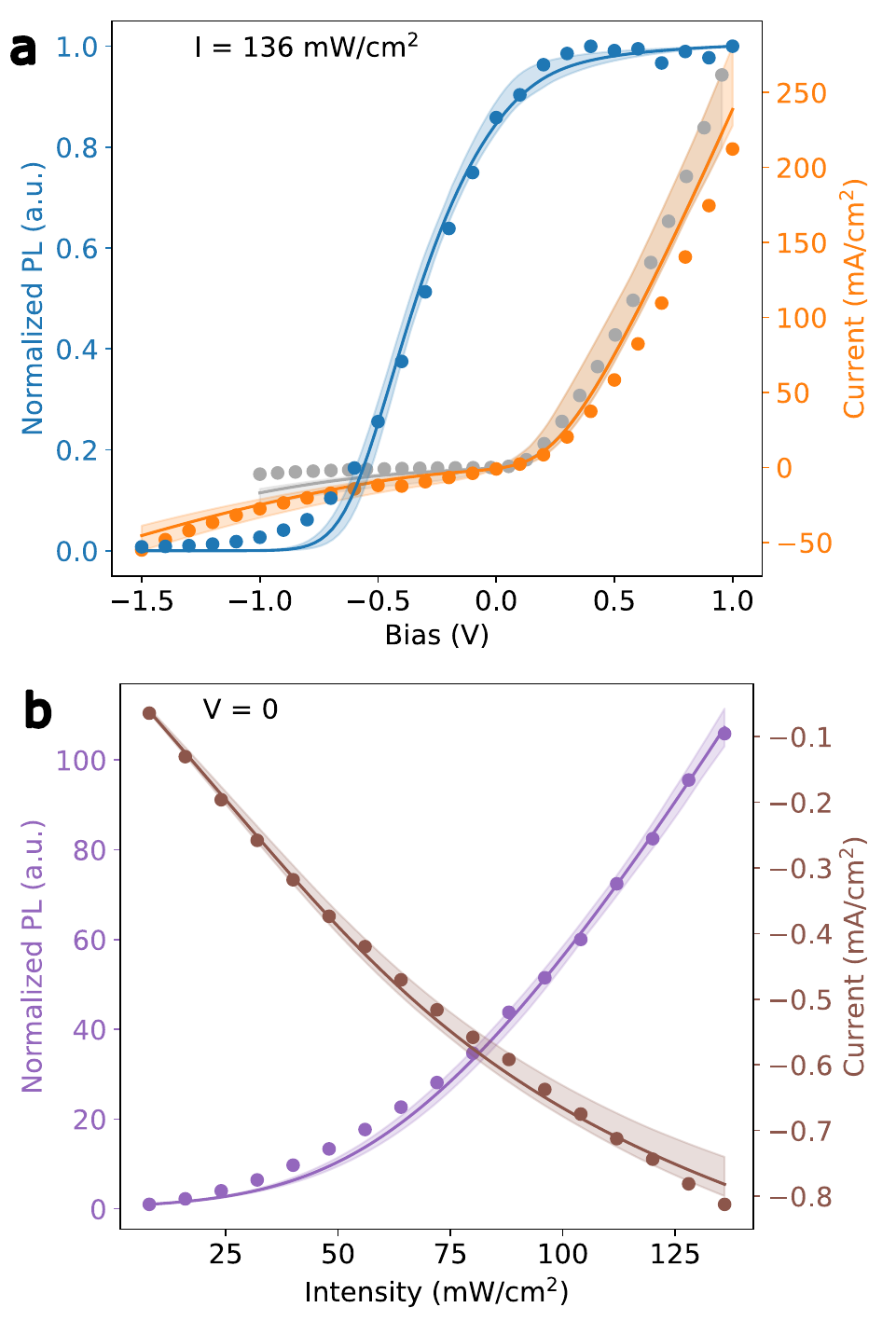}
\caption{\label{fig:pl_jv_fit}(a) Bias-dependent PL and current measurements
and fits. PL intensity is integrated over photon energy and normalized
to the maximum of the data set. (b) Intensity-dependent short-circuit
PL and current measurements and fits. PL intensity is integrated over
photon energy and normalized to the minimum of the data set. Experimental
data are in solid points. Best fits are shown with solid lines of
the same color. Shaded regions show 95\% confidence intervals.}
\end{figure}

Figure~\ref{fig:pl_jv_fit} shows the intensity- and voltage- dependent
measurements. The light JV (Fig.~\ref{fig:pl_jv_fit}(a), orange)
shows an S shape, similar to that observed in Ref.~\cite{esmaielpour_exploiting_2020}.
The bias-dependent integrated PL intensity (Fig.~\ref{fig:pl_jv_fit}(a),
blue) saturates at large voltage, unlike the expected result for an
LED. The intensity-dependent integrated PL intensity and short-circuit
current density $J_{\text{sc}}$ (Fig.~\ref{fig:pl_jv_fit}(b), purple
and brown, respectively) display nonlinear dependence on intensity.
These behaviors cannot be described by a standard two-diode model
for photovoltaics, in which radiative recombination grows exponentially
with applied voltage and PL intensity and $J_{\text{sc}}$ grow linearly
with illumination intensity. In the next section, we present an equivalent-circuit
model that is able to simultaneously fit these measurements.


\section{Equivalent circuit model}

Equivalent-circuit models are useful tools in describing light-emitting
diodes (LEDs) and photovoltaic cells. With a circuit model or lumped
parameter model, one can quickly predict current-voltage characteristics
as well as luminescence intensity under different bias and illumination
conditions. Before describing our proposed model, we start with a
brief review of the standard two-diode model for PV devices and a
previous model for S-shaped JV in a solar cell. 

\begin{figure}
\centering
\includegraphics[width=\columnwidth]{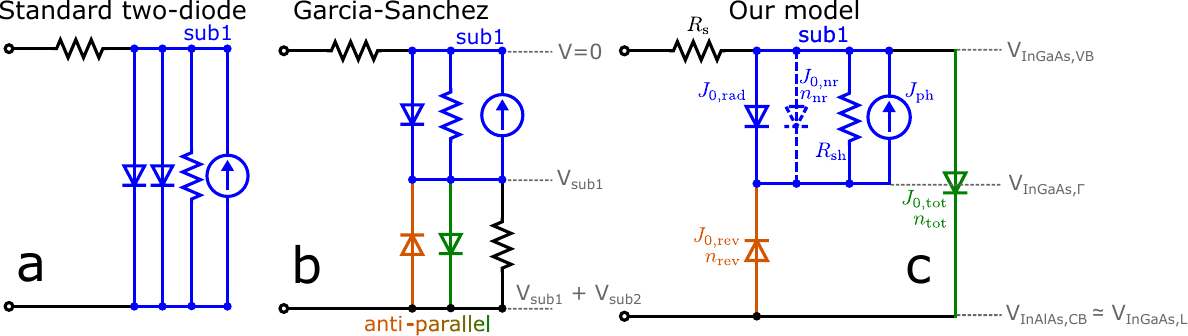}
\caption{\label{fig:circuits}(a) Standard two-diode model, (b) previous equivalent
circuit model for S-shaped JV, (c) our proposed model. In (c), the
top bus voltage can be interpreted as the quasi-Fermi level of InAlAs
VB at the p-type contact, $V_{\text{InAlAs, VB}}$, the middle bus
at the InGaAs $\Gamma$ valley, $V_{\text{InGaAs,}\Gamma}$, and the
bottom bus at the InAlAs CB at the n-type contact, $V_{\text{InAlAs, CB}}$.
With small potential barrier between InGaAs L valley and InAlAs CB,
we consider their quasi-Fermi level to be approximately equal, $V_{\text{InAlAs,CB}}=V_{\text{InGaAs, L}}$.}
\end{figure}

A two-diode model, shown in Fig.~\ref{fig:circuits}(a), can describe
most standard photovoltaic devices. The circuit contains two diodes
and a current source connected in parallel, complemented by series
and shunt resistors. Physically, the current source provides the photocurrent
$J_{\text{ph}}$, which is proportional to illumination intensity.
The series resistance $R_{\text{s}}$ accounts for both internal (due
to finite mobility) and external (at the contacts) resistances. The
shunt resistance $R_{\text{sh}}$ is large for good devices and is
small if there is a short between the p and n contacts. Large $R_{\text{s}}$
or small $R_{\text{sh}}$ reduce fill factor. The two diodes describe
radiative and non-radiative recombination effects. For each diode
$i$, the current density $J_{i}$ with voltage drop $V_{i}$ is
\begin{equation}
J_{i}=J_{0,i}\left(e^{qV_{i}/n_{i}kT}-1\right),
\end{equation}
where $J_{0,i}$ is the reverse-saturation current, $n_{i}$ is the
ideality factor, and $q$ is the elementary charge. Radiative recombination
produces photoluminescence, so the PL intensity $I_{\text{rad}}$
can be estimated as 
\begin{equation}
I_{\text{rad}}=S_{\text{pl}}J_{0,\text{rad}}e^{qV_{\text{rad}}/kT},\label{eq:diode-radiation}
\end{equation}
where $V_{\text{rad}}$ is the voltage across the radiative diode
and $S_{\text{pl}}$ is a constant dependent on device and detector
geometry. For radiative recombination, $n_{\text{rad}}=1$. For non-radiative
recombination, $n_{\text{nr}}$ depends on the dominant mechanism
and in practice can vary with voltage \cite{mcintosh_depletion-region_2000}.
If a voltage $V_{\text{sub}1}$ is applied across the blue part of
the circuit in Fig.~\ref{fig:circuits}(a) (subcircuit-1), the resulting
current is 
\begin{equation}
\begin{split}
J_{\text{sub1}}\left(V_{\text{sub1}}\right)={}&-J_{\text{ph}}+J_{0,\text{rad}}\left(e^{qV_{\text{sub1}}/kT}-1\right)\\
&+J_{0,\text{nr}}\left(e^{qV_{\text{sub1}}/n_{\text{nr}}kT}-1\right)+\frac{V_{\text{sub1}}}{R_{\text{sh}}}.
\end{split}
\end{equation}
We then include the series resistance to get the total $J(V)$. Since
the resistor is in series, the total current is $J_{\text{sub1}}$,
and $J=J_{\text{sub1}}=V_{\text{s}}/R_{\text{s}}$, and the total voltage
$V=V_{\text{s}}+V_{\text{sub1}}$, where $V_{s}$ is the voltage drop
across the series resistor. Therefore, $V_{\text{sub1}}$ is a function
of total voltage and current, $V_{\text{sub1}}(V,J)=V-JR_{\text{s}}$,
so the total current is implicitly defined, $J=J_{\text{sub1}}(V-JR_{\text{s}})$,
and can be solved numerically. By fitting the circuit parameters to
experimentally measured $J(V)$, one can estimate the strength and
dominating mechanism of nonradiative recombination in the device,
which affect device performance.

Not all solar cells can be modeled by the standard two-diode circuit.
In particular, this circuit cannot produce an S-shaped JV. The Garcia-Sanchez
model extends the two-diode model and adds a pair of anti-parallel
diodes in series with subcircuit-1, as illustrated in Fig.~\ref{fig:circuits}b
\cite{garcia-sanchez_lumped_2013}. Here the nonradiative diode in
subcircuit 1 is eliminated, because in fitting, the nonradiative diode
cannot be distinguished from the added forward diode. In the anti-parallel
pair, the reverse diode is responsible for the S shape, and the forward
diode allows the current to grow exponentially in forward bias. Despite
good agreement between the Garcia-Sanchez model and S-shaped JV measurements,
the model predicts an LED-like exponentially increasing PL in forward
bias, while our experimental frequency-integrated PL intensity saturates
in forward bias even as the current continues to grow. 
If we neglect the series resistor, the total voltage across the device is $V=V_{\text{sub1}}+V_{\text{sub2}}$,
and from Eq.~\ref{eq:diode-radiation}, the PL
intensity grows exponentially with $V_{\text{rad}}=V_{\text{sub1}}$. 
As the total applied bias $V$ increases, that voltage is shared between the subcircuits 
according to their differential conductances $G_i=d J_{\text{sub }i}/dV_{\text{sub }i}$, for $i=1,2$,
with the voltage increase going proportionally
to the subcircuit with smaller $G_i$. Under large forward biases, both 
subcircuits have currents dominated by their forward diodes, so both of them have 
differential conductances $G_i$ that increase exponentially with subcircuit voltage $V_{\text{sub }i}$. 
We then conclude that as $V$ increases,  $V_\text{sub1}$ must also increase; in the alternative case,
 where the entire marginal voltage drop is across subcircuit 2, the differential conductance of 
 subcircuit 2 would increase without bound, which is not consistent with it having
 the full voltage drop. 
Since $V_\text{sub1}$ increases with $V$, this model 
does not allow saturation of the PL intensity, regardless of the parameters chosen.

We modify the Garcia-Sanchez model so the S-shaped JV and forward-bias-saturated
PL can be simultaneously fitted by the equivalent circuit. Instead
of having the forward diode in parallel with the reverse diode, we
take it to be in parallel with the entire circuit, as shown in Fig.~\ref{fig:circuits}(c).
Subcircuit 2 is only the reverse diode, and $V_{\text{rev}}=V_{\text{sub2}}$.
In this case, the voltage across the forward diode is the sum of subcircuit
1 and subcircuit 2 voltages, $V_{\text{fwd}}^{\text{our model}}=V_{\text{sub1}}+V_{\text{sub2}}$.
We show the full $J(V)$ relation of our modified equivalent circuit
in the supplementary material, Section S1. In our circuit, the reverse diode still
produces the S shape. The total forward diode still allows diode-like
JV at large forward bias, but the PL intensity saturates because 
the reverse diode limits the current through the radiative diode.

We use our equivalent circuit model to perform a global fit of the
bias- and intensity-dependent PL as well as dark JV, illuminated JV
at 136 mW/cm$^{2}$, and intensity-dependent short-circuit current,
$J_{\text{sc}}$. We eliminate the nonradiative diode in subcircuit
1 since it cannot be distinguished from the forward diode when fitting.
For fitting our data points with various illumination intensities,
we recall that $J_{\text{ph}}$ is proportional to illumination intensity.
We fix the proportionality constant in our fits at $I_{\text{ref}}=136$
mW/cm$^{2}$. Below, in our AM1.5 JV fits, $I_{\text{ref}}=1$ sun.
The fits and 95\% confidence interval (CI) are shown in Fig.~\ref{fig:pl_jv_fit}.
Fit parameters and uncertainties are listed in the first column of
Table~\ref{tab:Circuit-parameters}. Uncertainties are 1$\sigma$,
estimated from the covariance matrix of the least squares fit. We
fit to the logarithm of $J_{0}$, $R_{s}$, and $R_{sh}$, producing
asymmetric uncertainties. Our fit captures all major experimental
features: saturating PL in forward bias, S-shaped JV, and nonlinear
PL and $J_{\text{sc}}$ as functions of illumination intensity. Our
model predicts an electroluminescence (EL) signal smaller than the
experimental noise floor of the bias- dependent PL signal, consistent
with experiments, in which we do not observe pure EL signals.

\begin{figure}
\centering
\includegraphics[width=\columnwidth]{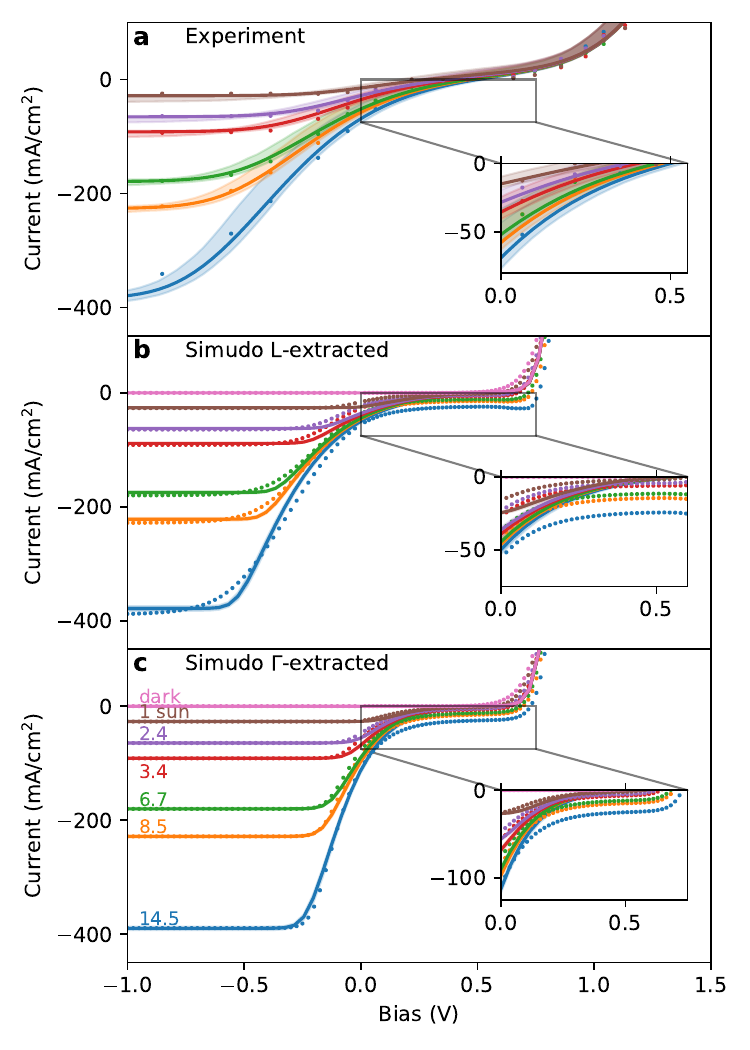}
\caption{\label{fig:jv-fit-all}Fit to JV curves from (a) experimental JV measurements
with AM1.5 illumination, digitized from \cite{esmaielpour_exploiting_2020},
(b) simulated, valley-scattered, as in Fig.~\ref{fig:device_band_diagrams}(b),
and (c) simulated $\Gamma$-extracted, as in Fig.~\ref{fig:device_band_diagrams}(c).
The solid lines are fits to the equivalent circuit model, with fit
parameters shown in Table 1, and the points are experimental or simulated
data. The faded areas are 95\% confidence intervals of the fits.}
\end{figure}

We also perform a global fit of the intensity-dependent JV under AM1.5
illumination digitized from Ref.~\cite{esmaielpour_exploiting_2020}.
In this fit, we include the nonradiative diode, which, unlike with
the PL dataset, improves fitting in the S-shape region. The fit results
are shown in Fig.~\ref{fig:jv-fit-all}(a). The fit parameters are
shown in Table~\ref{tab:Circuit-parameters}. Our fit well reproduces
the S shapes and the voltage knees where the S starts, which shift
to larger reverse bias with increasing intensity. 

\begin{table*}
\caption{Circuit parameters\label{tab:Circuit-parameters}}
{\renewcommand{\arraystretch}{1.5}%
\begin{tabular*}{\textwidth}{@{\extracolsep\fill}lllll@{}}
\toprule
\textbf{Parameter} & \textbf{PL\&JV} & \textbf{AM1.5 \cite{esmaielpour_exploiting_2020}} & \textbf{PDD Valley-scattered} & \textbf{PDD $\boldsymbol{\Gamma}$-extracted}\\
\midrule
$J_{0,\text{rad}}$ & $(4.16^{+0.31}_{-0.29})\times10^{-2}$ & $(1.02^{+18.6}_{-0.97})\times10^{-7}$ & $1.02\times10^{-7}${*} & $1.02\times10^{-7}${*}\\
$J_{0,\text{nr}}$ & - & $2.08^{+2.79}_{-1.19}$ & - & -\\
$n_{\text{nr}}$ & - & $4.36\pm0.94$ & - & -\\
$J_{0,\text{rev}}$ & $1.53^{+0.17}_{-0.16}$ & $13.3^{+6.5}_{-4.4}$ & $4.01^{+0.46}_{-0.41}$ & $1.08^{+0.17}_{-0.15}$\\
$n_{\text{rev}}$ & $12.0\pm1.0$ & $10.5\pm1.0$ & $8.31\pm0.26$ & $4.53\pm0.14$\\
$J_{0,\text{tot}}$ & $1.06^{+2.07}_{-0.70}$ & $(9.73^{+11.2}_{-5.40})\times10^{-2}$ & $(1.35^{+0.15}_{-0.14})\times10^{-11}$ & $(2.52^{+0.20}_{-0.18})\times10^{-11}$\\
$n_{\text{tot}}$ & $2.80\pm0.85$ & $6.09\pm0.70$ & 1{*} & 1{*}\\
$R_{\text{s}}$ & $(2.55^{+0.25}_{-0.23})\times10^{-3}$ & $(5.14^{+0.73}_{-0.64})\times10^{-4}$ & $(1.54^{+0.22}_{-0.20})\times10^{-4}$ & $(1.48^{+0.12}_{-0.11})\times10^{-4}$\\
$R_{\text{sh}}$ & $(1.28^{+0.11}_{-0.10})\times10^{-2}$ & - & - & -\\
\bottomrule
\end{tabular*}}
\begin{tablenotes}
\item[] Current densities are in mA/cm$^{2}$, resistances are in m$\Omega\cdot$cm$^{2}$.\quad {*}Fixed parameter in fitting
\end{tablenotes}
\end{table*}

\section{Physical meanings of circuit elements}

Having found an equivalent circuit that well describes the experimental
results, we now discuss the physical meanings of the circuit elements.
The blue subcircuit in Fig.~\ref{fig:circuits}(c) is the same as
the standard two-diode model, and the elements have their usual meanings.
The total forward diode shown in green in Fig.~\ref{fig:circuits}(c)
causes the diode-like JV in large forward bias. Such a diode could
represent a Schottky-like shunt or recombination in InAlAs. Ref.~\cite{breitenstein_shunt_2004}
reported a Schottky-like shunt with larger-than unity ideality factor
in a silicon solar cell, which can be caused by direct contact between
the n-type metallization and the p-type region. Reference~\cite{kanda_analysis_2016}
reports ITO sputtering damage on perovskite solar cells that behaves
as a total shunt diode. 

The reverse diode is key to the S-shaped JV, saturation of bias-dependent
PL, and the lack of electroluminescence. To interpret the physical
meaning of the reverse diode, we start with the physical meaning of
the buses in the circuit. Figure \ref{fig:circuits}(c) labels the
top bus $V_{\text{InGaAs,VB}}$ and the bottom bus as both $V_{\text{InAlAs,CB}}$
and $V_{\text{InGaAs,L}}$. The combined potential drop across the
blue sub-circuit and the reverse diode can be interpreted as the potential
difference between the InAlAs VB at the p-type contact and the InAlAs
CB at the n-type contact. In both valley-scattering and heterojunction
models (see Fig.~\ref{fig:device_band_diagrams}(b,c)), there is
no direct electrical connection from the InGaAs $\Gamma$ valley to
the contacts, so we consider the middle bus to represent the quasi-Fermi
level (QFL) of the InGaAs $\Gamma$ valley, $V_{\text{InGaAs,}\Gamma}$.
In the valley-scattering mode, there is only a small QFL difference
between the n-type InAlAs CB and InGaAs L valley, due to the small
energy barrier. Hence, we consider $V_{\text{InGaAs,L}}\simeq V_{\text{InAlAs, CB}}$.
Thus, the reverse diode represents physical processes between InGaAs
$\Gamma$ valley and InGaAs L valley, or equivalently, between InGaAs
$\Gamma$ valley and n-InAlAs CB. For each of these scenarios, we
propose a physical mechanism that can give rise to the reverse diode.

(i) $\Gamma$-extraction. The reverse diode can represent thermionic
emission at the heterojunction (HJ) barrier between the conduction
band (CB) of the n-InAlAs and the n-InGaAs layer. In this scenario,
we consider that electrons are primarily extracted from the $\Gamma$
valley to InAlAs CB, as indicated in Fig.~\ref{fig:device_band_diagrams}(c),
and the reverse diode represents the HJ barrier between InGaAs and
InAlAs, which is 0.74~eV. Ref.~\cite{kumar_current_1968} derived
the current-voltage expression under depletion approximation for an
nn isotype HJ, as with our n-InAlAs/n-InGaAs HJ. Ref.~\cite{kumar_current_1968}
shows that an nn-HJ in dark may be described by two opposing diodes
in series. Our blue sub-circuit includes two forward diodes for the
radiative and nonradiative processes, as well as a current source
representing the photocurrent, but the blue sub-circuit forms a forward-diode-like
behavior; adding in the reverse diode, our circuit is the illuminated
analogue to the expression in Ref.~\cite{kumar_current_1968}. Hence,
as described below, our equivalent circuit with the reverse diode
explains the heterojunctions in VPV well.

(ii) Valley scattering. The potential difference across the reverse
diode represents the QFL split between the populations of $\Gamma$
and L, with the L-valley carriers of the InGaAs sharing a QFL with
the $\Gamma$-valley conduction band carriers of InAlAs.

In scenario (i), there is no evidence of VS processes in the data.
In scenario (ii), VS is essential in describing the results, and electrical
extraction from the L valley of InGaAs is occurring.

\subsection*{Origins of the reverse diode}

We perform Poisson/drift-diffusion (PDD) studies of both scenarios
using Simudo \cite{dumitrescu_simudo_2020}. For scenario (ii), we
implement valley scattering using the method described in detail in
Ref.~\cite{xia_device-scale_2025}. In brief, we consider the net
valley scattering generation rate $g^{\text{VS}}$ from $\Gamma$
to L with rates extracted from spatially uniform Ensemble Monte Carlo
(EMC) simulations, provided by David Ferry. In order to obey detailed
balance at equilibrium, we take those EMC rates to be a function of
a quasi-electric field $\mathcal{E}_{k}$ in each valley $k$, where
$\mathcal{E}_{k}$ is the physical electric field when the system
is spatially uniform and is zero at equilibrium; we take $\mathcal{E}_{k}\equiv\nabla w_{k}/q$
where $w_{k}$ is the QFL of carriers in valley $k$. We take
\begin{equation}
g^{\text{VS}}=r_{\Gamma}\left(\mathcal{E}_{\Gamma}\right)u_{\Gamma}-r_{\text{L}}\left(\mathcal{E}_{\text{L}}\right)u_{\text{L}},\label{eq:net_g_vs}
\end{equation}
 where $u_{k}$ are carrier densities and the electric-field dependent
scattering rates $r_{k}$ are taken from the EMC study. At equilibrium,
$\mathcal{E}_{k}=0$, and the carrier populations in $\Gamma$ and
L valleys obey detailed balance, 
\begin{equation}
r_{\Gamma}\left(\mathcal{E}_{\Gamma}=0,w^{\text{eq}}\right)=r_{\text{L}}\left(0\right)\frac{u_{\text{L}}\left(w^{\text{eq}}\right)}{u_{\Gamma}\left(w^{\text{eq}}\right)},
\end{equation}
where $w^{\text{eq}}$ is the equilibrium Fermi level. In our PDD
model, we treat $r_{\text{L}}(0)$ as a free parameter and choose
it so that the JV curves qualitatively match the experiment data in
Fig.~\ref{fig:jv-fit-all}(a). The net valley scattering rate when
$\mathcal{E}_{\Gamma}=\mathcal{E}_{L}=0$ is 
\begin{equation}
g^{\text{VS}}(\mathcal{E}_{\Gamma}=\mathcal{E}_{\text{L}}=0)=u_{\text{L}}r_{\text{L}}(0)\left[\text{exp}\left(\frac{w_{\Gamma}-w_{\text{L}}}{kT}\right)-1\right],\label{eq:vs-diode-form}
\end{equation}
which is a diode-like form, though the factor of $u_{\text{L}}$
makes it not an ideal diode. Our results show that the device with
the same structure as in Ref.~\cite{esmaielpour_exploiting_2020}
operates at small $\mathcal{E}_{L}$, so the valley scattering process
is diode like, with the diode open when $w_{\Gamma}>w_{\text{L}}$,
which describes a reverse diode. Therefore, we expect that our equivalent
circuit with the reverse diode fits scenario (ii) well. 

In experimental devices, extraction from $\Gamma$ in scenario (i)
and from L in scenario (ii) can happen in parallel. Scenarios (i)
and (ii) are two limiting cases. If both cases give reverse-diode-like
behavior, we expect combinations also to be reverse-diode like.

We simulate both scenarios using Simudo \cite{dumitrescu_simudo_2020},
with material parameters in the supplementary, Section~S2.
We consider an InAlAs/InGaAs/InAlAs device with optical absorption
and associated radiative recombination but no nonradiative processes.
Scenario (i), which we call $\Gamma$-extracted, is a standard two-band
semiconductor model without VS, using a thermionic HJ boundary condition
at the InGaAs/InAlAs interfaces \cite{yang_numerical_1993}, with
only one carrier population per band. The band diagram at equilibrium
is shown in Fig.~\ref{fig:device_band_diagrams}(c). In scenario
(ii), which we call valley scattered, we model InGaAs with three independent
quasi-equilibrated carrier populations, representing carriers in the
valence band and the $\Gamma$ and L valleys of the conduction
band \cite{xia_device-scale_2025}. Only the L valley population is
electrically coupled to the InAlAs CB, so electrons generated in $\Gamma$
need to valley scatter to L before being extracted to InAlAs. The
equilibrium band diagram of scenario (ii) is shown in Fig.~\ref{fig:device_band_diagrams}(b).
In both scenarios, we adjust one simulation parameter (HJ band alignment
for (i), $r_{\text{L}}(0)$ for (ii)), to find JV curves that are
qualitatively similar to the experimental JV in Fig.~\ref{fig:jv-fit-all}(a).
Our PDD model treats the carrier populations in each band as being
in quasi-equilibrium and therefore does not include hot-carrier effects
beyond excess occupancy of the L valley. 

We fit the resulting JV curves to our equivalent-circuit model. Since
the simulations do not include nonradiative processes, we eliminate
the non-radiative diode. In our simulation, we do not model any Schottky-diode-like
shunt resistance, such as that caused by imperfect fabrication, so
the total diode in our simulation has to be the recombination in InAlAs.
Hence, we set $n_{\text{tot}}=1$, assuming that the total diode describes
the radiative recombination in InAlAs. Without the non-radiative diode
and with $n_{\text{tot}}=1$, $J_{0,\text{rad}}$ and $J_{\text{0,tot}}$
covary in our fits, so we fix $J_{\text{0,rad}}$ to be constant and
the same as the fitted $J_{0,\text{rad}}$ in the fit of the experimental
AM1.5 JV. Results are shown in Fig.~\ref{fig:jv-fit-all}(b,c) with
fitted parameters and uncertainties in Table~\ref{tab:Circuit-parameters}.
In the valley-scattered case, the reverse diode is best fit with ideality
factor $n_{\text{rev}}=8.31$, while Eq.~\ref{eq:vs-diode-form}
appears to show an ideality factor of 1. The $u_{\text{L}}$ factor
in Eq.~\ref{eq:vs-diode-form} makes it not exactly a diode form
and can shift the ideality factor, as can having $\mathcal{E}$ not
equal to zero in either band. The equivalent circuit fits to the PDD
results are good in both scenarios (i) and (ii), indicating that both
HJ and VS can produce the reverse diode in our circuit, and we cannot
easily distinguish the mechanisms.

\section{Discussion}

As a summary, our equivalent-circuit model can simultaneously describe
PL and electrical measurements of VPV devices. Our equivalent circuit
also fits well to illumination-dependent S-shape JV curves. The S
shape is caused by a reverse diode, which can be explained by two
possible physical processes, (i) heterojunction barrier between InGaAs
$\Gamma$ and InAlAs CB, (ii) the valley-scattering process itself.

\begin{figure}
\centering
\includegraphics[width=\columnwidth]{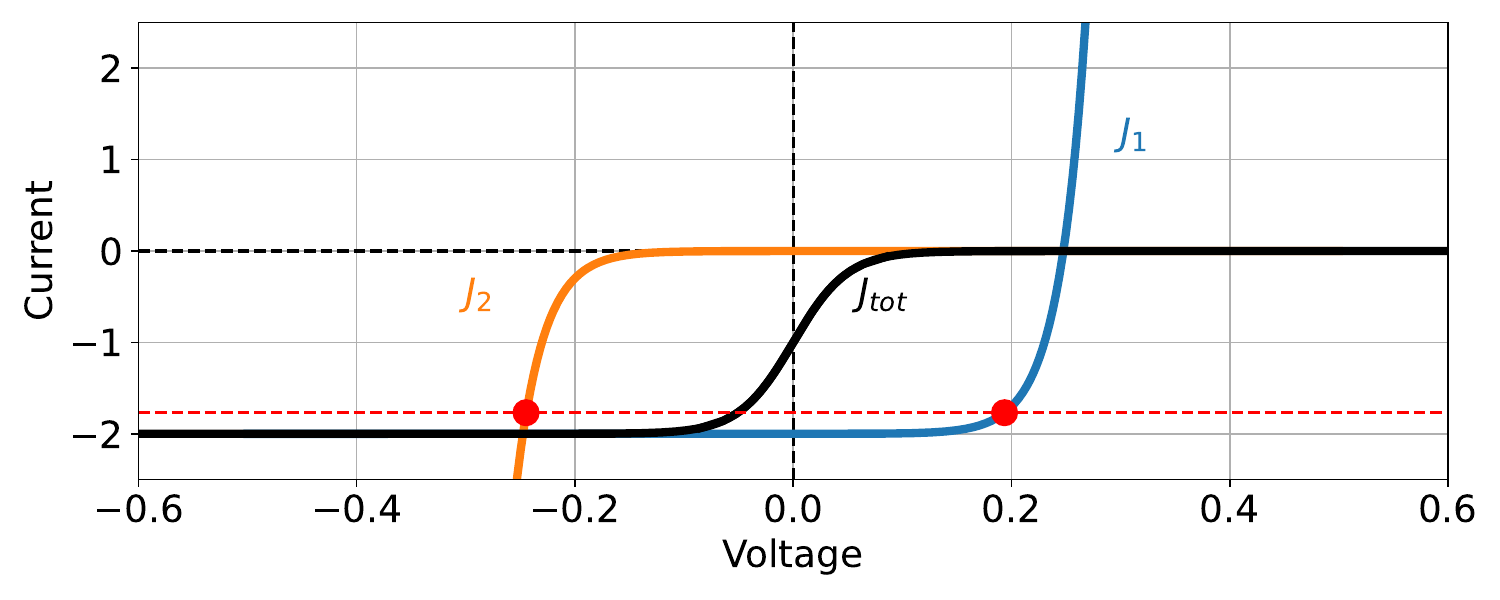}
\caption{\label{fig:rev-diode-demo}Current-voltage characteristics of a forward
diode (blue), a reverse diode (orange), and the total current when
they are in series. The red dots and horizontal line indicate the
current where the forward diode is at its maximum power point. For
any point in the power-generating (fourth) quadrant on the forward
diode JV, the current-matching condition when these diodes are in
series requires that the reverse diode be in negative voltage. Hence
the entire device's voltage must be reduced by the presence of the
reverse diode compared to a circuit with only the forward diode.}
\end{figure}

As long as our equivalent-circuit model describes the VPV devices,
the reverse diode prevents these devices from achieving high efficiency.
As demonstrated in Fig.~\ref{fig:rev-diode-demo}, when connecting
the reverse diode in series with a conventional PV equivalent circuit,
the reverse diode can only have a negative voltage when the conventional
circuit is in the power-generating quadrant. Therefore, the reverse
diode always hurts the voltage and the efficiency of the device compared
to an alternative structure that does not have the reverse diode.
This effect can be seen in the total current, which displays an S
shape, where at all values of current, the total voltage is smaller
than the forward-diode voltage.

If the reverse diode represents VS, then we do see the VS effect in
this device, but this type of VS does not help with efficiency. For
the VS effect to help with the device efficiency, VS must not behave
like a reverse diode. Eq.~\ref{eq:vs-diode-form} shows a diode form
when the quasi-electric fields are zero. In our PDD modeling of the
VPV device of the same structure as in Ref.~\cite{esmaielpour_exploiting_2020},
we find that the device indeed operates in the small-field regime
\cite{xia_device-scale_2025}. If there were a device architecture
that could support large $\mathcal{E}_{\Gamma}$ at the operating
voltage, then from Eq.~\ref{eq:net_g_vs}, the net VS generation
to L would be large and thus could improve device efficiency. Alternatively,
eliminating the reverse diode and achieving high efficiency may require
high carrier temperatures in addition to the separately quasi-equilibrated
$\Gamma$ and L populations in our PDD model \cite{xia_device-scale_2025}.
If the reverse diode originates in the HJ barrier, then we can explain
the full PL and electrical data without reference to VS. In that case,
we do not see evidence of VS, and the reverse diode has to be eliminated
before VS effects can be seen and potentially be useful.

\bmsubsection*{Acknowledgments}
We thank Dave Ferry for sharing the ensemble Monte Carlo valley scattering rates.
We acknowledge support from Natural Sciences and Engineering Research Council of Canada [NSERC CREATE TOP-SET (Award 497981) and RGPIN-2019-06559] and M.P.N. recognizes the support of the UNSW Scientia Program and an Australian Research Council (ARC) DECRA Fellowship (DE230100382).

\bibliographystyle{vancouver-authoryear}
\bibliography{reference}

\onecolumn
\setcounter{section}{0}\setcounter{equation}{0}\setcounter{figure}{0}\setcounter{table}{0}
\renewcommand{\thesection}{S\arabic{section}}
\renewcommand{\theequation}{S\arabic{equation}}
\renewcommand{\thefigure}{S\arabic{figure}}
\renewcommand{\thetable}{S\arabic{table}}

\begin{center}
{\Large\bfseries Supplementary material for: Combined photoluminescence and electrical characterization of valley photovoltaic devices explained with an equivalent circuit\par}
\end{center}
\section{Derivation of current-voltage relation}

Here, we derive the current-voltage relation $J(V)$ for our equivalent
circuit, shown in Fig.~4c of the main text. The
sub1 current takes its usual form:

\begin{equation}
J_{\text{sub1}}\left(V_{\text{sub1}}\right)=-J_{\text{ph}}+J_{0,\text{rad}}\left(e^{qV_{\text{sub1}}/kT}-1\right)+J_{0,\text{nr}}\left(e^{qV_{\text{sub1}}/n_{\text{nr}}kT}-1\right)+\frac{V_{\text{sub1}}}{R_{\text{sh}}}.\label{eq:one}
\end{equation}
The reverse-diode current is equal to the sub1 currents, since they
are series connected, giving,
\begin{equation}
J_{\text{rev}}(V_{\text{rev}})=J_{\text{sub1}}=J_{0,\text{rev}}\left(e^{qV_{\text{rev}}/n_{\text{rev}}kT}-1\right).\label{eq:two}
\end{equation}
Then we can calculate $V_{\text{sub2}}=V_{\text{rev}}$ as
\begin{equation}
V_{\text{rev}}=\frac{n_{\text{rev}}kT}{q}\text{ln}\left[\frac{J_{\text{sub1}}(V_{\text{sub1}})}{J_{0,\text{rev}}}+1\right].\label{eq:three}
\end{equation}
Then the total voltage without series resistance, which we call the
internal voltage $V_{\text{int}}$, is the sum of $V_{\text{sub1}}$
and $V_{\text{rev}}$,
\begin{equation}
V_{\text{int}}(V_{\text{sub1}})=V_{\text{sub1}}+\frac{n_{\text{rev}}kT}{q}\text{ln}\left[\frac{J_{\text{sub1}}(V_{\text{sub1}})}{J_{0,\text{rev}}}+1\right].\label{eq:four}
\end{equation}

The forward diode is in parallel to the series-connected sub1 and
reverse diode. The current of the forward diode is then
\begin{equation}
J_{\text{fwd}}(V_{\text{sub1}})=J_{0,\text{fwd}}\left[e^{qV_{\text{int}}(V_{\text{sub1}})/n_{\text{fwd}}kT}-1\right].\label{eq:five}
\end{equation}
The total current along both parallel branches is then
\begin{align}
J(V_{\text{sub1}}) & =J_{\text{sub1}}(V_{\text{sub1}})+J_{\text{fwd}}(V_{\text{sub1}})\nonumber \\
 & =-J_{\text{ph}}+J_{0,\text{rad}}\left(e^{qV_{\text{sub1}}/kT}-1\right)+J_{0,\text{nr}}\left(e^{qV_{\text{sub1}}/n_{\text{nr}}kT}-1\right)+\frac{V_{\text{sub1}}}{R_{\text{sh}}}\label{eq:total_current}\\
 & \,\,\,\,\,\,\,\,+J_{0,\text{fwd}}\left\{ \exp\left[\frac{qV_{\text{sub1}}}{n_{\text{fwd}}kT}+\frac{n_{\text{rev}}}{n_{\text{fwd}}}\text{ln}\left(\frac{-J_{\text{ph}}+J_{0,\text{rad}}\left(e^{qV_{\text{sub1}}/kT}-1\right)+J_{0,\text{nr}}\left(e^{qV_{\text{sub1}}/n_{\text{nr}}kT}-1\right)+\frac{V_{\text{sub1}}}{R_{\text{sh}}}}{J_{0,\text{rev}}}+1\right)\right]-1\right\} .\nonumber 
\end{align}
The voltage drop across the series resistance is 
\[
V_{\text{s}}=J(V_{\text{sub1}})R_{\text{s}}.
\]
The total voltage is then 
\begin{align}
V(V_{\text{sub1}}) & =V_{\text{int}}+V_{\text{s}}\nonumber \\
 & =V_{\text{sub1}}+\frac{n_{\text{rev}}kT}{q}\text{ln}\left[\frac{J_{\text{sub1}}(V_{\text{sub1}})}{J_{0,\text{rev}}}+1\right]+J(V_{\text{sub1}})R_{\text{s}}\nonumber \\
 & =V_{\text{sub1}}+\frac{n_{\text{rev}}kT}{q}\text{ln}\left[\frac{-J_{\text{ph}}+J_{0,\text{rad}}\left(e^{qV_{\text{sub1}}/kT}-1\right)+J_{0,\text{nr}}\left(e^{qV_{\text{sub1}}/n_{\text{nr}}kT}-1\right)+\frac{V_{\text{sub1}}}{R_{\text{sh}}}}{J_{0,\text{rev}}}+1\right]\nonumber \\
 & \,\,\,\,\,\,\,\,+R_{\text{s}}\Big(\label{eq:total_voltage}\\
 & \,\,\,\,\,\,\,\,\,\,\,\,\,\,\,\,\,\,-J_{\text{ph}}+J_{0,\text{rad}}\left(e^{qV_{\text{sub1}}/kT}-1\right)+J_{0,\text{nr}}\left(e^{qV_{\text{sub1}}/n_{\text{nr}}kT}-1\right)+\frac{V_{\text{sub1}}}{R_{\text{sh}}}\nonumber \\
 & \,\,\,\,\,\,\,\,\,\,\,\,\,\,\,\,\,\,+J_{0,\text{fwd}}\left\{ \exp\left[\frac{qV_{\text{sub1}}}{n_{\text{fwd}}kT}+\frac{n_{\text{rev}}}{n_{\text{fwd}}}\text{ln}\left(\frac{-J_{\text{ph}}+J_{0,\text{rad}}\left(e^{qV_{\text{sub1}}/kT}-1\right)+J_{0,\text{nr}}\left(e^{qV_{\text{sub1}}/n_{\text{nr}}kT}-1\right)+\frac{V_{\text{sub1}}}{R_{\text{sh}}}}{J_{0,\text{rev}}}+1\right)\right]-1\right\} \nonumber \\
 & \,\,\,\,\,\,\,\,\,\,\,\,\,\,\,\,\,\,\Big).\nonumber 
\end{align}
Combining Eqs.~\ref{eq:total_current} and \ref{eq:total_voltage},
we have the $J(V)$ relationship implicitly defined through $V_{\text{sub1}}$.

\section{PDD simulation parameters}

Table \ref{tab:SI-PDD-parameters} shows the material parameters
used in the PDD simulations. $\mu_{i}$ is the mobility of carriers
in band $i$, $N_{i}$ is the effective density of states of band
$i$, and $E_{i}$ is the band extremum (maximum or minimum) of band
$i$.

\begin{table}[!htb]
\centering
\caption{PDD simulation parameters}\label{tab:SI-PDD-parameters}

\centering{}%
\begin{tabular}{lcc}
 &  & \tabularnewline
\hline 
\hline 
Parameter & PDD Valley-scattered & PDD $\Gamma$-extracted\tabularnewline
\hline 
$\mu_{\text{L,InGaAs }}$ & \multicolumn{2}{c}{444~cm$^{2}$/V/s}\tabularnewline
$\mu_{\Gamma\text{,InGaAs }}$ & \multicolumn{2}{c}{$1.39\times10^{4}$~cm$^{2}$/V/s}\tabularnewline
$\mu_{\text{VB,InGaAs }}$ & \multicolumn{2}{c}{490~cm$^{2}$/V/s}\tabularnewline
$\mu_{\text{CB,InAlAs }}$ & \multicolumn{2}{c}{517~cm$^{2}$/V/s}\tabularnewline
$\mu_{\text{VB,InAlAs }}$ & \multicolumn{2}{c}{136~cm$^{2}$/V/s}\tabularnewline
$N_{\Gamma\text{,InGaAs }}$ & \multicolumn{2}{c}{$2.10\times10^{17}$~cm$^{-3}$}\tabularnewline
$N_{\text{L,InGaAs }}$ & \multicolumn{2}{c}{$6.67\times10^{19}$~cm$^{-3}$}\tabularnewline
$N_{\text{VB,InGaAs }}$ & \multicolumn{2}{c}{$7.37\times10^{18}$~cm$^{-3}$}\tabularnewline
$N_{\text{CB,InAlAs }}$ & \multicolumn{2}{c}{$4.85\times10^{17}$~cm$^{-3}$}\tabularnewline
$N_{\text{VB,InAlAs }}$ & \multicolumn{2}{c}{$1.10\times10^{19}$~cm$^{-3}$}\tabularnewline
$E_{\Gamma\text{,InGaAs }}$ & \multicolumn{2}{c}{0.72 eV}\tabularnewline
$E_{\text{L,InGaAs }}$ & \multicolumn{2}{c}{1.25 eV}\tabularnewline
$E_{\text{VB,InGaAs }}$ & \multicolumn{2}{c}{0 eV}\tabularnewline
$E_{\text{CB,InAlAs }}$ & 1.31 eV & 1.59 eV\tabularnewline
$E_{\text{VB,InAlAs }}$ & $-0.14$ eV & 0.14 eV\tabularnewline
\hline 
\hline 
 &  & \tabularnewline
\end{tabular}
\end{table}

\end{document}